\documentclass[aps,prd,reprint,superscriptaddress,amsmath,amssymb,nofootinbib,longbibliography,floatfix]{revtex4-2}

\usepackage{graphicx}
\usepackage{dcolumn}
\usepackage[utf8]{inputenc}
\usepackage{epstopdf}
\usepackage{xcolor}
\usepackage[unicode=true,colorlinks,linkcolor=blue,anchorcolor=blue,urlcolor=blue,citecolor=blue,breaklinks=true]{hyperref}
\usepackage{subfigure}
\usepackage{makecell}

\newcommand{\pararrow}{\mathord{\buildrel{\lower3pt\hbox{$\scriptscriptstyle\leftrightarrow$}}\over {\partial}}} 
\newcommand{\pararrowk}[1]{\mathord{\buildrel{\lower3pt\hbox{$\scriptscriptstyle\leftrightarrow$}}\over {\partial}\hspace*{-0.18em}{}^#1}\hspace*{-0.18em} \,} 

\usepackage{slashed}
\usepackage{booktabs}
\usepackage{physics}
\usepackage{bm}
\usepackage{mathrsfs}
\newcommand{\qfnu}{\affiliation{College of Physics and Engineering, Qufu Normal University, Qufu 273165, China}}

\newcommand{\ucas}{\affiliation{School of Nuclear Science and Technology, University of Chinese Academy of Sciences, Beijing 101408, China}}

\newcommand{\scnt}{\affiliation{Southern Center for Nuclear-Science Theory (SCNT), Institute of Modern Physics, Chinese Academy of Sciences, Huizhou 516000, China}}

\newcommand{\imp}{\affiliation{State Key Laboratory of Heavy Ion Science and Technology, Institute of Modern Physics, Chinese Academy of Sciences, Lanzhou 730000, China}}

\newcommand{\tju}{\affiliation{Center for Joint Quantum Studies and Department of Physics, School of Science, Tianjin University, Tianjin 300350, China}}

\begin{document}
\title{Final-state rescattering mechanism of the $\Delta(1232)^{++}$ production in $\Lambda^+_c \to K^- \pi^+ p$ decay}

    \author{Zu-Xin Cai} \qfnu \imp
	\author{Cheng Chen} \imp \ucas
    \author{Si-Wei Liu} \imp \ucas
    \author{Xiang Wei} \imp \ucas
    \author{Gang Li}~\email{gli@qfnu.edu.cn} \qfnu 
    \author{Xiao-Hai Liu}~\email{xiaohai.liu@tju.edu.cn} \tju
    \author{Ju-Jun Xie}~\email{xiejujun@impcas.ac.cn} \imp \ucas \scnt

\date{\today}

\begin{abstract}

We investigate the production of the $\Delta(1232)^{++}$ resonance in the charmed baryon weak decay $\Lambda^+_c \to K^- \pi^+ p$, focusing on the $\pi^+ p$ final-state rescattering mechanism. The direct $W^+$ exchange diagram is expected to be suppressed, hence we adopt the $W^+$ internal emission process $\Lambda^+_c \to p \bar K^{*0}(892)$ followed by the subsequent decay $\bar{K}^{*0} \to K^- \pi^+$ as the dominant source of the final state particles. The $\Delta(1232)^{++}$ resonance is then generated via $\pi^+ p$ rescattering within a triangle loop mechanism. Our calculations incorporate both the tree-level $\bar K^{*0}(892)$ and the dynamically generated $\bar{K}^*_0(700)$ state arising from the $S$-wave $K \pi$ final state interaction. We find that our theoretical results can reproduce the bump and peak structures in the $K^- \pi^+$ invariant mass distributions for the $\bar{K}^*_0(700)$ and $\bar{K}^{*0}(892)$, respectively. Meanwhile, the peak for the $\Delta(1232)^{++}$ in the $\pi^+ p$ invariant mass distributions is also well described. The $\Delta(1232)^{++}$ signal naturally emerges from rescattering effects, and adopting the pole parameters of $\Delta(1232)$ resonance yields an improved description of the experimental data. In addition, we obtain a branching fraction ratio $\mathcal{B}[\Lambda_c^+ \to \Delta(1232)^{++} K^-] / \mathcal{B}[\Lambda_c^+ \to p \bar{K}^{*0}(892)] \approx 0.5$, which is lower than the experimentally measured value. This discrepancy suggests that interference effects are likely significant in this decay process. Future high-precision measurements will further verify the proposed rescattering mechanism.

\end{abstract}

\maketitle

\section{Introduction}
\label{sec:Introduction}

The multi-body nonleptonic decays of charmed baryons are useful processes for studying the nature of hadronic resonances~\cite{Cheng:2015iom,Oset:2016lyh,Cheng:2018hwl}. The $\Lambda_c^+$ weak decays have been used to investigate the $\Lambda(1405)$ and $\Lambda(1670)$ resonances~\cite{Miyahara:2015cja,Xie:2016evi,Wang:2022nac,Zhang:2024jby,Duan:2024okk}, the $\Sigma(1385)$ and a possible $\Sigma(1380)$ state~\cite{Xie:2017xwx,Xie:2019zzb,Lyu:2024qgc}, the $N(1535)$ resonance~\cite{Xie:2017erh,Xie:2020jlz,Li:2024rqb,Li:2026lbo}, the $f_0(980)$ and $a_0(980)$ resonances~\cite{Wang:2020pem,Feng:2020jvp}, and the $\Xi^*(1690)$ resonance~\cite{Liu:2023jwo}. These studies clearly demonstrate that the $\Lambda_c^+$ weak decays provide a valuable source for investigating low-lying hadronic states. Great progress has been made in experimental measurements of $\Lambda^+_c$ hadronic weak decays~\cite{Belle:2013jfq,BESIII:2015bjk,LHCb:2017xtf,Belle:2018gcs,Belle:2020xku,BESIII:2020kzc,BESIII:2022udq,BESIII:2024xny,BESIII:2025rda,BESIII:2026mpt,BESIII:2026qbp}. The $\Lambda^+_c \to \Lambda \pi^+ \eta$ decay was measured by the BESIII Collaboration~\cite{BESIII:2018qyg,BESIII:2024mbf}, where the $a_0(980)^+$ production was observed for the first time, and evidence for the potential pentaquark state $\Sigma(1380)^+$ was found in the $\pi^+ \Lambda$ system. In Ref.~\cite{Belle:2022pwd}, evidence for the production of $N^*(1535)$ resonance with $N(1535)^+ \to p \eta$ was found in the $\Lambda^+_c \to K^0_S p \eta$ decay by the Belle Collaboration. These rich experimental results have promoted extensive theoretical interest in studying low-lying hadronic states and related topics.

Among the three-body $\Lambda_c^+$ hadronic weak decays, $\Lambda^+_c \to K^- \pi^+ p$ has attracted sustained experimental and theoretical interest~\cite{E791:1999ajq,Liu:2019dqc,Ahn:2019rdr,LHCb:2022sck} due to its large branching fraction~\cite{Belle:2013jfq,BESIII:2015bjk,Wei:2022kem,ParticleDataGroup:2024cfk}, and its complex intermediate resonant structures~\cite{E791:1999ajq,Ahn:2019rdr,Marangotto:2020ead}. A recent amplitude analysis of $\Lambda_c^+ \to p K^- \pi^+$ decay performed by the LHCb Collaboration reveals a pronounced diagonal band associated with the $\Delta(1232)^{++}$ resonance in the Dalitz plot constructed from $m^2_{K^-\pi^+}$ versus $m^2_{p K^-}$. A clear peak of the $\Delta(1232)^{++}$ resonance is also found in the $\pi^+ p$ invariant mass spectrum around $M^2_{\pi^+ p} \sim 1.5~\mathrm{GeV}^2$~\cite{LHCb:2022sck}. Moreover, the $\bar K^{*0}(892)$ resonance yields a substantial contribution, manifesting as a clear peak peak in the $K^-\pi^+$ invariant mass distribution. Contributions from excited hyperon $\Lambda$ states are also present in the $K^- p$ subsystem.

The $\Delta(1232)$ resonance, carrying quantum numbers $I(J^P)=3/2(3/2^+)$, was initially observed by Anderson, Fermi, and their colleagues via $\pi p$ scattering measurements~\cite{Anderson:1952nw}. As the lightest excited baryon resonance, it has Breit-Wigner mass of $1232 \pm 2$~MeV and a decay width of $117 \pm 3$~MeV, while its pole mass and pole width are $1210 \pm 1$~MeV and $100 \pm 2$~MeV, respectively. The dominant decay channel of $\Delta(1232)$ proceeds into the $\pi N$ final states. Within the constituent quark model and under flavor $SU(3)$ symmetry, it belongs to the baryon decuplet. The $\Delta(1232)$ is widely adopted as an intermediate resonance produced via the $s$-channel process~\cite{Huang:2011as,Dai:2025hvo}. Experimentally, the LEPS2/BGOegg Collaboration observed a clear $\Delta(1232)$ peak in the $\pi^0 p$ invariant mass spectrum, which further confirms its important role in relevant reactions~\cite{LEPS2BGOegg:2023ssr}. Moreover, the $\Delta(1232)$ resonance has also been interpreted as a composite $\pi N$ molecular state. Theoretical studies indicate that its wave function contains a substantial $\pi N$ component~\cite{Aceti:2014ala,Sekihara:2015aba,Sekihara:2015gvw}.

In the $\Lambda_c^+ \to K^- \pi^+ p$ decay, if the $\Delta(1232)^{++}$ were produced as an intermediate state in $s$-channel, at the quark level the process would proceed through a color-suppressed $W^+$ exchange diagram ($c+d \to s+u$). Such contributions are generally small~\cite{Cheng:2010vk,Cheng:2015iom,Miyahara:2016yyh,Xie:2014tma} and therefore cannot account for the large signal observed experimentally. Thus, we adopt the $W^+$ internal emission mechanism, assuming that the $ud$ diquark ($I=0$) in the $\Lambda_c^+$ acts as a spectator. This leads to the decay chain $\Lambda_c^+ \to p\bar{K}^{*0}(892) \to K^- \pi^+ p$, from which the $\Delta(1232)^{++}$ is subsequently produced through the final-state rescattering of $\pi^+ p$. Within this framework, we study the productions of $\bar K^{*0}(892)$ and $\Delta(1232)^{++}$ resonances in the $\Lambda_c^+ \to K^- \pi^+ p$ decay using the effective Lagrangian approach. To better describe the event concentration in the low-energy region of the $K^- \pi^+$ invariant mass distribution, we also include the contribution of the $\bar{K}^*_0(700)$ resonance with $I(J^P) = 1/2(0^+)$ (denoted as $\kappa$)~\footnote{In what follows, we will use $\kappa$ to denote $K^*_0(700)$ since the scalar meson $K^*_0(700)$ is also known as $\kappa$.}, which is dynamically generated from the $S$-wave pseudoscalar meson-pseudoscalar meson interaction~\cite{Pelaez:2020gnd,Albaladejo:2025lhn}. We present the $K^- \pi^+$ and $p\pi^+$ invariant mass distributions, as well as the Dalitz plot of $M^2_{K^-\pi^+}$ and $M^2_{p\pi^+}$, for the $\Lambda^+_c \to K^- \pi^+ p$ decay. It is shown that the bump and peak structures in the $K^- \pi^+$ invariant mass distributions for the $\kappa$ and $\bar{K}^{*0}(892)$ resonances can be well reproduced. Meanwhile, the peak for the $\Delta(1232)^{++}$ in the $\pi^+ p$ invariant mass distributions can be also well described.

The paper is organized as follows. In Sec.~\ref{sec:Formalism}, we present the theoretical formalism for the production of $\kappa$, $\bar{K}^{*0}(892)$, and $\Delta(1232)^{++}$ in the $\Lambda^+_c \to K^- \pi^+ p$ decay. In Sec.~\ref{sec:Results} the numerical results and discussions are presented, and finally a brief summary is given in Sec.~\ref{sec:summary}.

\section{Formalism} \label{sec:Formalism}

In this section, we present the theoretical decay formalism for the process $\Lambda^+_c \to K^- \pi^+ p$. In Sec.~\ref{subsec:FSI}, we describe the $S$-wave $K\pi$ final-state interaction, from which the scalar meson $\kappa$ is dynamically generated. In Sec.~\ref{subsec:Formalism_K_Delta}, we introduce the tree-level $\bar K^{*0}(892)$ contribution and the $\Delta(1232)^{++}$ contribution generated through $\pi^+ p$ rescattering. Finally, in Sec.~\ref{subsec:Formalism_IMD}, the formulas used to calculate the $K^-\pi^+$ and $p\pi^+$ invariant mass distributions for the $\Lambda^+_c \to K^- \pi^+ p$ decay are given.

\subsection{Dynamically generated scalar meson $\kappa$ from the $S$-wave $K \pi$ final-state interaction} \label{subsec:FSI}

\begin{figure}[htbp]
    \centering
    \includegraphics[scale=0.42]{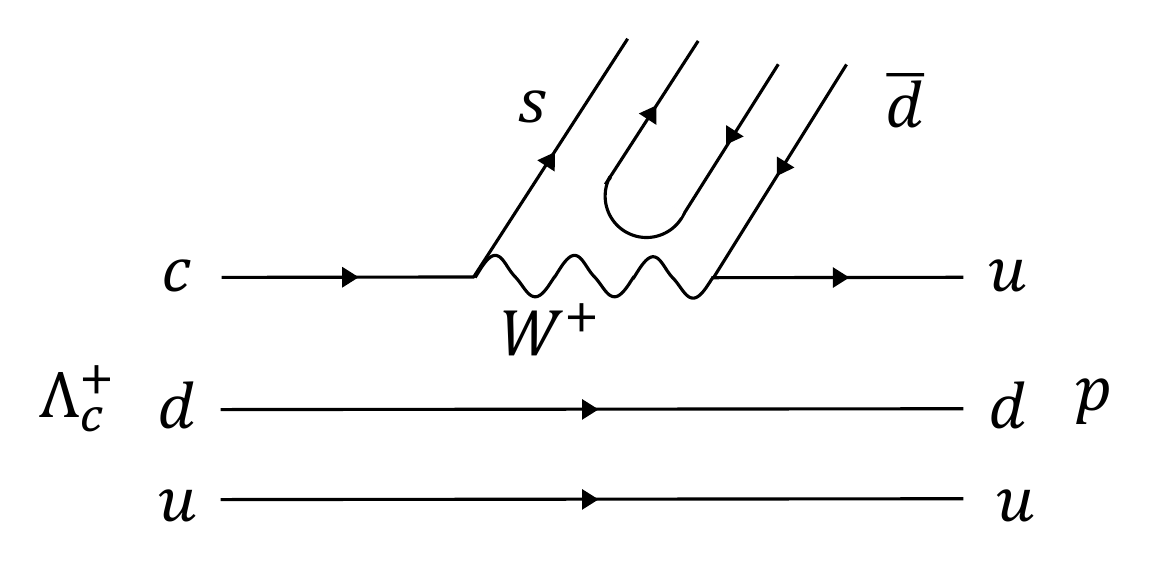}
    \caption{Quark level diagram for the charm quark in the $\Lambda_c^+$ decaying into a strange quark and a $\bar{d}u$ pair. The proton is formed by the up quark and the spectator $ud$ diquark in the $\Lambda_c^+$.} 
    \label{fig:Quark_Level}
\end{figure}

\begin{figure}[htbp]
    \centering
    \includegraphics[scale=0.3]{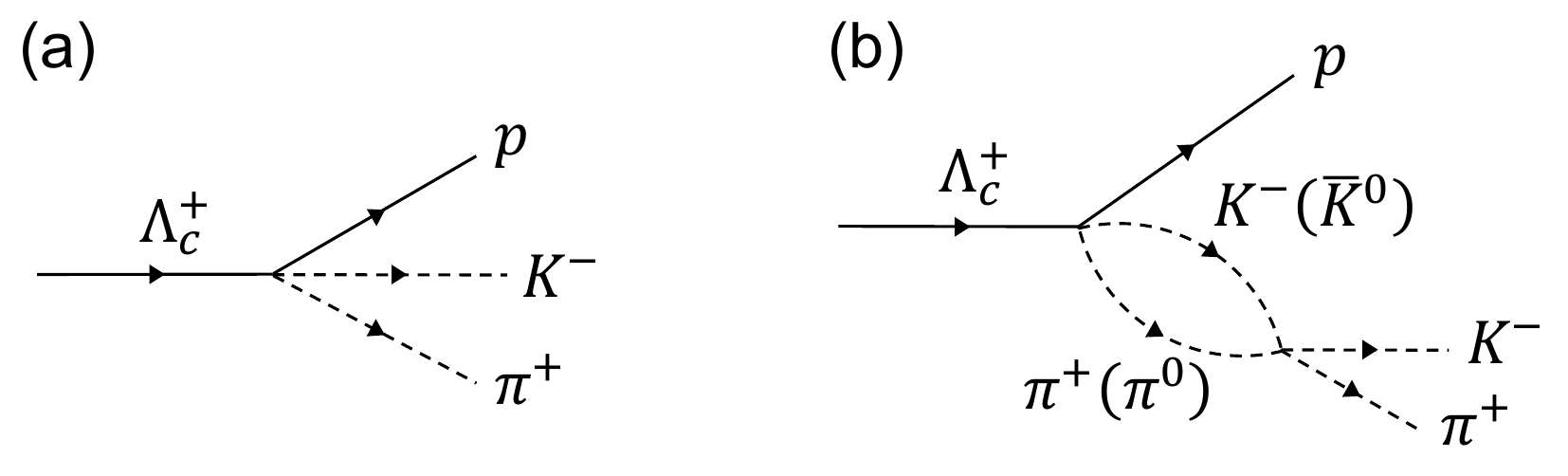}
    \caption{Diagrams for the $\Lambda^+_c \to K^- \pi^+ p$ decay: (a) direct $K^- \pi^+ p$ vertex at tree level; (b) final-state interaction of the $K^- \pi^+$.}
    \label{fig:Hadron_Level}
\end{figure}

Following previous works~\cite{Xie:2017erh,Wang:2020pem,Feng:2020jvp}, we consider the $W^+$ internal emission mechanism for the $\Lambda^+_c \to K^- \pi^+ p$ process, which is illustrated in Fig.~\ref{fig:Quark_Level}. In this mechanism, the $c$ quark in $\Lambda_c^+$ decays into an $s$ quark and a $W^+$ boson, followed by the $W^+$ boson decays into an $u\bar{d}$ pair. Under the spectator assumption, the $ud$ diquark in the $\Lambda_c^+$ combines with the $u$ quark from the weak decay to form the final proton. The remaining $s\bar d$ pair hadronizes into two pseudoscalar mesons by combining with a $\bar{q}q$ pair with the quantum numbers of the vacuum. This hadronization process can be written as
\begin{align}
    \Lambda_c^+&\Rightarrow \frac{1}{\sqrt{2}}c(ud-du) \Rightarrow \frac{1}{\sqrt{2}}s W^+ (ud-du)\nonumber\\
    &\Rightarrow \frac{1}{\sqrt{2}}s\bar{d}u(ud-du) \Rightarrow s(\bar{u}u+\bar{d}d+\bar{s}s)\bar{d}\,p\nonumber\\
    &= \left(\sum_i M_{3i}M_{i2}\right)p \, .
    \label{eq:quark_level}
\end{align}
Here, $M$ denotes the $SU(3)$ pseudoscalar meson matrix
\begin{align}
    M &= \begin{pmatrix}
        \frac{\pi^0}{\sqrt{2}}+\frac{\eta}{\sqrt{3}} & \pi^+ & K^+ \\
        \pi^- & -\frac{\pi^0}{\sqrt{2}}+\frac{\eta}{\sqrt{3}} & K^0  \\
        K^- & \bar{K}^0 & -\frac{\eta}{\sqrt{3}}
    \end{pmatrix},
\end{align}
where we have ignored the $\eta'$ term as done in Refs.~\cite{Khemchandani:2016ftn,Miyahara:2016yyh}, because of its relatively large mass. Substituting the pseudoscalar meson matrix into Eq.~(\ref{eq:quark_level}) yields
\begin{align}
    \Lambda_c^+\Rightarrow  \left(K^- \pi^+ - \frac{1}{\sqrt{2}} \bar{K}^0 \pi^0\right)p .
    \label{eq:MM_channel}
\end{align}
The first term in Eq.~(\ref{eq:MM_channel}) indicates that the $K^- \pi^+ p$ final state can be produced directly as depicted in Fig.~\ref{fig:Hadron_Level} (a). In addition, the $K^- \pi^+$ pair can also be dynamically generated from the $S$-wave interaction involving $K^-\pi^+$ and $\bar{K}^0 \pi^0$, as presented in Fig.~\ref{fig:Hadron_Level} (b).

Then, the decay amplitude associated with the above hadronization process can be written as
\begin{align}
    {\mathcal{M}}_\kappa &= V_p \Big[ 1+G_{K^- \pi^+}(M_{K^- \pi^+}) T_{K^- \pi^+ \to K^- \pi^+}(M_{K^- \pi^+}) \nonumber \\
    &- \frac{1}{\sqrt{2}} G_{\bar{K}^0 \pi^0}(M_{K^- \pi^+}) T_{\bar{K}^0 \pi^0 \to K^- \pi^+}(M_{K^- \pi^+}) \Big],
    \label{eq:amp_kappa}
\end{align}
where $V_p$ is an overall normalization constant including the weak decay contribution. We assume $V_p$ to be constant and determine it from the experimental data on the $\kappa$ production. $M_{K^- \pi^+}$ denotes the invariant mass of the $K^- \pi^+$ system. Besides, $G$ and $T$ are the meson-meson loop function and the two-body scattering amplitudes, respectively, both of which depend on the invariant variable $s=M^2_{K^-\pi^+}$. The loop function $G$ is regularized with a three-momentum cutoff,
\begin{align}
	16\pi^2sG(s) & = \sigma\left(\arctan\frac{s+m_1^2-m_2^2}{\sigma\lambda_1} + \right.  \nonumber \\
    & \hspace{-1.68cm} \left. \arctan\frac{s-m_1^2+m_2^2}{\sigma\lambda_2}\right)  -\left[ (s+m_1^2-m_2^2)\times \right. \nonumber \\
    & \hspace{-1.68cm} \left. \ln\frac{(1+\lambda_1)q_{\max}}{m_1} +(s-m_1^2+m_2^2)\ln\frac{(1+\lambda_2)q_{\max}}{m_2}\right], \label{eq:loopfunction}
\end{align}
where $\sigma$, $\lambda_1$, and $\lambda_2$ are defined as
\begin{eqnarray}
	\sigma &=& \left[-\left(s-(m_1+m_2)^2\right)\left(s-(m_1-m_2)^2\right)\right]^{1/2} \, , \\
  \lambda_{1,2} &=& \sqrt{1+\frac{m_{1,2}^2}{q^2_{\rm max}}}\, . 
\end{eqnarray} 
Here, $m_1$ and $m_2$ are the masses of the two mesons in the loop. In this work, we take the cutoff parameter $q_\mathrm{max} = 600$~MeV, as used in Ref.~\cite{Toledo:2020zxj}.

The two-body scattering amplitude $T$ is obtained by solving the Bethe-Salpeter equation within the chiral unitary approach \cite{Oller:1997ti,Oller:1998hw,Oller:1997ng},
\begin{align}
    T = [1-VG]^{-1}V\,.
\end{align}
As done in Refs.~\cite{Gamermann:2006nm,Toledo:2020zxj}, the symmetric matrix elements of $V$ can be written as
\begin{align}
    V_{11} &=-\frac{1}{6f^2} \Big( \frac{3}{2}s- \frac{3}{2s}(m_\pi^2-m_K^2)^2 \Big),\\
    V_{12} &=\frac{1}{2 \sqrt{2}f^2} \Big( \frac{3}{2}s- m_\pi^2-m_K^2-  \frac{(m_\pi^2-m_K^2)^2}{2s} \Big),\\
    V_{22} &=-\frac{1}{4f^2} \Big( -\frac{s}{2}+ m_\pi^2+m_K^2-  \frac{(m_\pi^2-m_K^2)^2}{2s} \Big), \\
    V_{13} &=\frac{1}{2\sqrt{6}f^2} \Big( \frac{3}{2}s- \frac{7}{6}m_\pi^2-\frac{1}{2}m_\eta^2-\frac{1}{3} m_K^2 \nonumber \\
    &+ \frac{3}{2s}(m_\pi^2-m_K^2)(m_\eta^2-m_K^2) \Big), \\
    V_{23} &=-\frac{1}{4\sqrt{3}f^2} \Big( \frac{3}{2}s- \frac{7}{6}m_\pi^2-\frac{1}{2}m_\eta^2-\frac{1}{3}m_K^2 \nonumber \\
    &+ \frac{3}{2s}(m_\pi^2-m_K^2)(m_\eta^2-m_K^2) \Big), \\
    V_{33} &=-\frac{1}{4f^2} \Big(- \frac{3}{2}s - \frac{2}{3}m_\pi^2+m_\eta^2+ 3m_K^2\nonumber \\
    &- \frac{3}{2s}(m_\eta^2-m_K^2)^2 \Big).
\end{align}
Here, the subscripts $1$, $2$, and $3$ correspond to the channels $\pi^+ K^-$, $\pi^0 \bar K^0$, and $\eta \bar K^0$, respectively. The pion decay constant is taken as $f=93$~MeV. The isospin averaged meson masses are $m_\pi=138.04$~MeV, $m_K=494.99$~MeV, and $m_\eta=547.86$~MeV\cite{ParticleDataGroup:2024cfk}. Within this coupled-channel $S$-wave interaction, the $\kappa$ resonance is dynamically generated and mainly contributes to the low-energy region of the $K^-\pi^+$ invariant mass spectrum.

\begin{table*}[htbp]
\renewcommand\arraystretch{1.5}
    \caption{Spin parities, masses, total widths, branching ratios, and extracted coupling constants for the $\Lambda_c^+$, $\bar K^{*0}(892)$, and $\Delta(1232)^{++}$. The couplings are obtained from Eqs.~(\ref{eq:g_LamcK892p}), (\ref{eq:g_K892piK}), and (\ref{eq:g_Delpip}). For the $\Delta(1232)^{++}$, Model~I uses the pole parameters, while Model~II uses the Breit-Wigner parameters~\cite{ParticleDataGroup:2024cfk}.}
    \label{table:Parameters}
    \begin{ruledtabular}
        \begin{tabular}{cccccc}
        State & Spin-parity($J^P$) & Mass (MeV) & Width (MeV) & $\mathcal{B}$ & Coupling \\
            \colrule
            $\Lambda_c^+$ & $1/2^+$ & $2286.46$ & $3.25\times 10^{-9}$ & $(1.41 \pm 0.07)\%$ & $g_{\Lambda^+_c p \bar K^{*0}}=3.96 \times 10^{-7}$ \\
            $\bar K^{*0}(892)$ & $1^-$ & $895.55$ & $47.30$ & $(99.754 \pm 0.021)\%$ & $g_{\bar K^{*0} \pi^+ K^-}=4.40$ \\
            $\Delta(1232)^{++}$ (Model I) & $3/2^+$ & 1210 & 100 & 99.4\% & $g_{\Delta^{++}\pi^+ p}=2.27$ \\
            $\Delta(1232)^{++}$ (Model II) & $3/2^+$ & 1232 & 117 & 99.4\% & $g_{\Delta^{++}\pi^+ p}=2.14$
        \end{tabular}
    \end{ruledtabular}
\end{table*}

\subsection{Tree-level $\bar K^{*0}(892)$ production and the $\Delta(1232)^{++}$ production via rescattering mechanism} \label{subsec:Formalism_K_Delta}

\begin{figure}[htbp]
    \centering
    \includegraphics[scale=0.28]{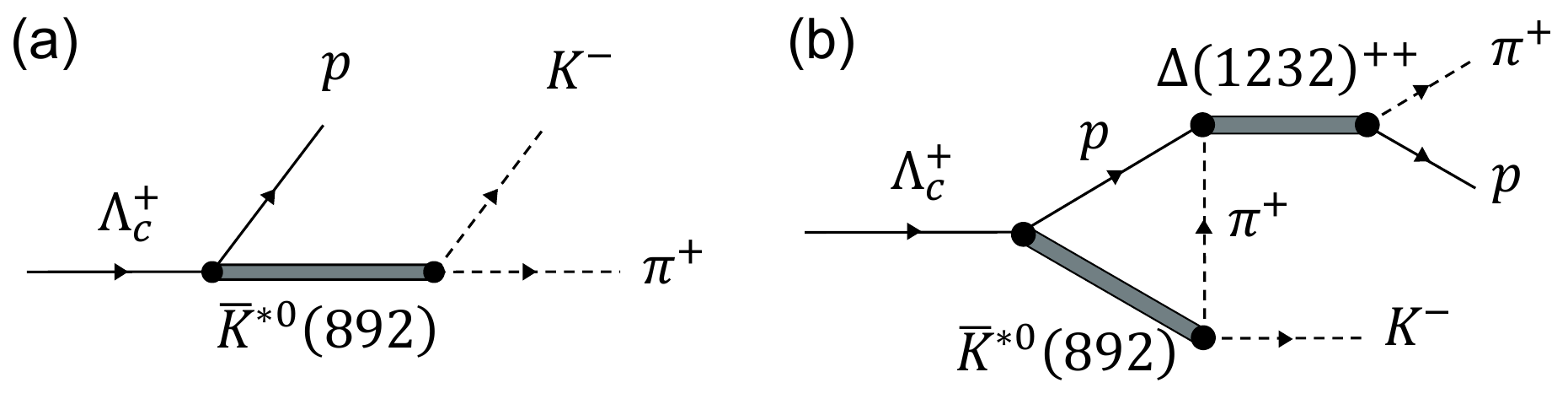}
    \caption{Mechanism for the production of $\bar{K}^{*0}(892)$ via tree diagram (a) and the production of $\Delta(1232)^{++}$ via triangle loop diagram (b) for the process $\Lambda_c^+(p_0)\to \pi^+(p_1) p(p_2) K^-(p_3)$. The intermediate particles in the triangle loop diagram are $\bar{K}^{*0}(q_1,m_1)$, $p(q_2,m_2)$, and $\pi^+ (q_3,m_3)$, where $q_i$ and $m_i$ denote their four-momenta and masses, respectively.}
    \label{fig:Feynman}
\end{figure}

In Fig.~\ref{fig:Feynman}, we show the tree level diagram and triangle diagram for the production of $\bar{K}^{*0}(892)$ and $\Delta(1232)^{++}$, respectively, in the $\Lambda_c^+ \to p K^- \pi^+$ decay. In Fig.~\ref{fig:Feynman}(a), the $\bar K^{*0}(892)$ resonance is formed from the $s\bar{d}$ pair, and the $u$ quark originated from the weak decay and the $ud$ diquark in the $\Lambda^+_c$ form the proton corresponding to the mechanism of Fig.~\ref{fig:Quark_Level}. After this hadronization, the $\bar{K}^{*0}$ then decays into $K^- \pi^+$ in the final state. For the $\Delta(1232)^{++}$ production, we consider the triangle diagram shown in Fig.~\ref{fig:Feynman}(b), in which the $\Delta(1232)^{++}$ is produced through the rescattering of $\pi^+ p \to \Delta(1232)^{++}$.

To evaluate the decay amplitude corresponding to the Feynman diagrams shown in Fig.~\ref{fig:Feynman}, we adopt the following relevant effective Lagrangians as used in previous studies~\cite{Ahn:2019rdr,Lee:2024vyq,Chen:2025bgq,Dai:2025hvo}
\begin{align}
    \mathcal{L}_{\Lambda_c^+ p \bar{K}^{*0}} &= -g_{\Lambda_c^+ p \bar{K}^{*0}}  \bar \Lambda_c^+   (g^\mathrm{PV}_{\Lambda_c^+ p \bar K^{*0}}\gamma_5 + g^\mathrm{PC}_{\Lambda_c^+ p \bar K^{*0}} ) \nonumber \\ 
    & \times \gamma^{\mu} {\bar{K}}^{*0}_{\mu} p  + \mathrm{h.c.}\,, \label{eq:VBB} \\
    \mathcal{L}_{\bar{K}^{*0} \pi^+ K^-} &= i g_{\bar{K}^{*0} \pi^+ K^-} \bar{K}^{*0}_\mu ( \pi^- \partial^\mu K^{-}  - K^{-} \partial^\mu \pi^-) \nonumber \\
    & + \mathrm{h.c.}\,, \label{eq:PPV} \\
    \mathcal{L}_{\Delta^{++} \pi^+ p} &= -\frac{i g_{\Delta^{++} \pi^+ p}}{m_{\pi^+}} \bar{\Delta}^{++}_\mu (\partial^\mu \pi^+) p + \mathrm{h.c.}\,. \label{eq:PBB'} 
\end{align}
Here, $g^\mathrm{PV}_{\Lambda_c^+ p \bar K^{*0}}$ and $g^\mathrm{PC}_{\Lambda_c^+ p \bar K^{*0}}$ denote the parity-violating (PV) and parity-conserving (PC) weak couplings, which are associated with different partial wave contributions. For simplicity, we set $g^\mathrm{PV}_{\Lambda_c^+ p \bar K^{*0}}=g^\mathrm{PC}_{\Lambda_c^+ p \bar K^{*0}}=1$ and absorb this common coupling into an effective coupling constant $g_{\Lambda_c^+ p \bar K^{*0}}$~\cite{Xie:2017xwx,Liu:2019dqc,Xie:2017erh}. The coupling constants $g_{\Lambda_c^+ p \bar K^{*0}}$, $g_{\Delta^{++} \pi^+ p}$, and $g_{\bar K^{*0} \pi^+ K^-}$ are determined from the corresponding partial decay widths, 
\begin{align}
    \Gamma_{\Lambda_c^+ \to p \bar K^{*0}}&= \frac{{g}_{\Lambda_c^+ p \bar K^{*0}}^{2}|{\vec{p}}_p|}{2\pi M_{\Lambda_c^+} m_{\bar K^{*0}}^2}  [E_p(M_{\Lambda_c^+}^2+m_p^2+2 m_{\bar K^{*0}}^2) \nonumber\\
    & -2M_{\Lambda_c^+} m_p^2] ~,
    \label{eq:g_LamcK892p}\\
    \Gamma_{\Delta^{++} \to \pi^+ p}&=\frac{{g}_{\Delta^{++} \pi^+ p}^{2} (E_p+m_p) {|{\vec{p}}_p|}^3}{12\pi m_\Delta m_\pi^2}~,
    \label{eq:g_Delpip} \\
    \Gamma_{\bar K^{*0} \to \pi^+ K^-}&=\frac{2{g}_{\bar K^{*0} \pi^+ K^-}^{2} {|{\vec{p}}_K|}^3}{3\pi m_{\bar K^{*0}}^2}~,
    \label{eq:g_K892piK}
\end{align}
where $|\vec{p}_p|$ and $|\vec{p}_K|$ are the three momenta in the rest frame of the corresponding decaying particle, and $E_p = \sqrt{|\vec{p}_p|^2 + m_p^2}$. These partial decay widths can be easily obtained from the total decay width and the corresponding branching fractions (${\cal B}$). Using the masses and partial widths shown in Table~\ref{table:Parameters}, the corresponding coupling constants are obtained, which are also listed in Table~\ref{table:Parameters}. To determine whether $\Delta(1232)^{++}$ is produced through the interaction of $\pi^+ p$ or exists as a simple Breit-Wigner resonance, we employed two models. In Model~I, we set the mass and width of $\Delta(1232)$ to its pole mass and width as quoted in the Particle Data Group~\cite{ParticleDataGroup:2024cfk}, while in Model~II, they are set to the Breit-Wigner mass and width~\cite{ParticleDataGroup:2024cfk}. 

With the effective Lagrangians given above, one can easily obtain the invariant decay amplitudes corresponding to Figs.~\ref{fig:Feynman}~(a)~and~\ref{fig:Feynman}~(b) as
\begin{align}
    {\mathcal{M}}_\mathrm{a} &= - g_{\Lambda_c^+ p \bar K^{*0}} g_{\bar K^{*0} \pi^+ K^-}\bar{u}(p_2)( \gamma_5+1) \gamma_\mu  \nonumber\\
    &\times G^{\mu \nu }_1 (p_1+p_3,m_{\bar K^{*0}},\Gamma_{\bar K^{*0}}) (p_{3}-p_{1})_\nu u(p_0)\, ,
    \label{eq:ampa} \\
    {\mathcal{M}}_\mathrm{b} &= \frac{g_{\Delta^{++}\pi^+ p}^2 g_{\Lambda_c^+ p \bar K^{*0}} g_{\bar K^{*0} \pi^+ K^-}}{m_3^2}\bar{u}(p_2) p_{1\mu} \nonumber\\ 
    &\times G_{3/2}^{\mu\nu}(p_{1}+p_2,m_\Delta,\Gamma_\Delta)\int \frac{\mathrm{d}^4 q_1}{(2\pi)^4} q_{3\nu} G_{1/2}(q_2,m_2,\epsilon) \nonumber\\
    &\times (\gamma_5+1) \gamma_\alpha G^{\alpha\beta}_1(q_1,m_1,\Gamma_{\bar K^{*0}})(p_3-q_3)_\beta \nonumber\\
    &\times G_0(q_3,m_3,\epsilon) u(p_0) F(q_3^2,m_3^2) \, ,
    \label{eq:ampb}
\end{align}
where $\epsilon$ is an infinitesimal quantity. The $m_{\Delta}$ and $\Gamma_\Delta$ stand for the mass and width of the $\Delta(1232)^{++}$ resonance, respectively. The propagators appearing in above equations are written in the unified form
\begin{align}
    G_J(q,m,\Gamma)=
    \frac{i\,\mathcal P_J(q,m)}
    {q^2-m^2+i m\Gamma}\,,
\end{align}
where $J=0,1,1/2,3/2$ denotes the spin of the intermediate particle. The numerator structures are
\begin{align}
    \mathcal P_0&=1\,,\qquad\qquad~
    \mathcal P_1^{\mu\nu}=-g^{\mu\nu}+\frac{q^\mu q^\nu}{m^2}\,, \nonumber\\
    \mathcal P_{1/2}&=\slashed q+m\,,\qquad
    \mathcal P_{3/2}^{\mu\nu}=(\slashed q+m)P_{3/2}^{\mu\nu}(q)\,, \notag
\end{align}
with
\begin{align}
    P_{3/2}^{\mu\nu}(q)=
    -g^{\mu\nu}
    +\frac{1}{3}\gamma^\mu\gamma^\nu
    +\frac{\slashed q}{3q^2}
    \left(\gamma^\mu q^\nu-q^\mu\gamma^\nu\right)
    +\frac{2q^\mu q^\nu}{3q^2}. \notag
\end{align}

For the triangle loop diagram, ${\cal M}_b$, a form factor $F(q^2,m^2)$ is introduced to account for off-shell effects and the internal structure of the exchanged particle~\cite{Gortchakov:1995im,Tornqvist:1993ng,Cheng:2004ru,Li:1996yn,Locher:1993cc,Shklyar:2005xg,Feuster:1997pq}. It also suppresses the ultraviolet behavior of the loop integral. In this work, we adopt the monopole form factor, as used in Refs.~\cite{Brockmann:1990cn,Machleidt:1987hj}:
\begin{align}\label{eq:ff1}
    F(q^2,m^2)=\frac{\Lambda^2-m^2}{\Lambda^2-q^2}\,,
\end{align}
where $q$ and $m$ are the four-momentum and mass of the exchanged $\pi^+$ meson, respectively. The $\Lambda$ is a cutoff parameter, whose value will be determined by comparing the theoretical calculations with the experimental measurements.

\subsection{Invariant mass distributions for the $\Lambda^+_c \to K^- \pi^+  p$ decay}\label{subsec:Formalism_IMD}

Combining the $S$-wave $K \pi$ final-state interaction, the tree-level $\bar{K}^{*0}$ production, and the triangle loop diagram, the decay amplitude of the process $\Lambda^+_c \to K^- \pi^+ p$ is constructed from ${\mathcal{M}}_\kappa$, ${\mathcal{M}}_\mathrm{a}$, and ${\mathcal{M}}_\mathrm{b}$. The first component is ${\mathcal{M}}_\kappa$, which describes the hadronization followed by the $S$-wave $K\pi$ interaction within the chiral unitary approach. The other two components, ${\mathcal{M}}_\mathrm{a}$ and ${\mathcal{M}}_\mathrm{b}$, are evaluated using the effective Lagrangian approach for the tree-level and triangle loop diagrams, respectively. In these decay amplitudes, the $\bar K^{*0}(892)$ acts as the intermediate particle, while the $\Delta(1232)^{++}$ is produced through $\pi^+ p$ final-state rescattering. Therefore, the total amplitude squared can be written as~\footnote{The interference between ${\cal M}_\kappa$ and ${\cal M}_a + {\cal M}_b$ is neglected. Since the $K^- \pi^+$ interaction corresponds to an $S$-wave component in one amplitude and $P$-wave component in the other, their interference is expected to be sufficiently small and can be safely omitted.}
\begin{align}\label{eq:amp_total}
    |\overline{\mathcal{M}}|^2 = C\left( |{\mathcal{M}}_\kappa|^2+\frac{1}{2}|{\mathcal{M}}_\mathrm{a}+{\mathcal{M}}_\mathrm{b}|^2 \right),
\end{align}
where the factor $1/2$ accounts for the average over the spins of the initial $\Lambda_c^+$. The factor $C$ is an overall normalization factor which is introduced to scale the theoretical invariant mass distributions to match the experimental measurements of the selected candidates. Subsequently, the double differential width of the $\Lambda^+_c \to K^- \pi^+ p$ decay is given by~\cite{ParticleDataGroup:2024cfk}
\begin{equation}\label{eq:dGamma}
	\frac{d^2 \Gamma}{d M_{K^- \pi^+}^2d M_{\pi^+p}^2}=\frac{1}{(2\pi)^3}\frac{1}{32M_{\Lambda_c^+}^3} |\overline{\mathcal{M}}|^2\,.
\end{equation}
Then the invariant mass squared distributions $d \Gamma/d M_{K^- \pi^+}^2$ and $d \Gamma/d M_{\pi^+p}^2$ can be obtained by integrating Eq.~(\ref{eq:dGamma}) over the other invariant-mass variable within the physical phase space. For a given value of $M_{\pi^+p}^2$, the kinematically allowed range of $M_{K^-\pi^+}^2$ is 
\begin{equation}
\begin{aligned}
    (M_{\pi^+ K^-}^2)_{\text{min}}&=(E_{\pi^+}+E_{K^-})^2 \\
    &-(\sqrt{E_{\pi^+}^2-m_{\pi^+}^2}+\sqrt{E_{K^-}^2-m_{K^-}^2})^2\,, \\
    (M_{\pi^+ K^-}^2)_{\text{max}}&=(E_{\pi^+}+E_{K^-})^2  \\
    &-(\sqrt{E_{\pi^+}^2-m_{\pi^+}^2}-\sqrt{E_{K^-}^2-m_{K^-}^2})^2\,, \notag
\end{aligned} 
\end{equation}
where $E_{\pi^+}$ and $E_{K^-}$ are the energies of particles $\pi^+$ and $K^-$ in the $\pi^+p$ rest frame, respectively, and are given by
\begin{equation}
\begin{aligned}
    E_{\pi^+}&=\frac{M_{\pi^+p}^2+M_{\pi^+}^2-M_{p}^2}{2M_{\pi^+p}}\,,\\
    E_{K^-}&=\frac{M_{\Lambda_c^+}^2-M_{\pi^+p}^2-M_{K^-}^2}{2M_{\pi^+p}}\,. \notag
\end{aligned} 
\end{equation}

\section{Numerical results} \label{sec:Results}

With the above formalism, we then calculate the invariant mass distributions of the process $\Lambda^+_c \to K^- \pi^+ p$. There are three free parameters to be obtained by fitting to the experimental data on the bump structure of $\kappa$, the $\bar{K}^{*0}$ peak, and the $\Delta^{++}$ peak: (1). $V_P$ for the weak and hadronization strength related to the $S$-wave $K\pi$ interaction of Fig.~\ref{fig:Hadron_Level} to produce the $\kappa$ state; (2). the global factor $C$ to match the theoretical calculations with the experimental measurements; (3). the cut off parameter $\Lambda$ for the exchanged $\pi^+$ meson in the triangle loop diagram as shown in Fig.~\ref{fig:Feynman} (b).

\begin{figure}[htbp]
    \centering
    \includegraphics[scale=0.45]{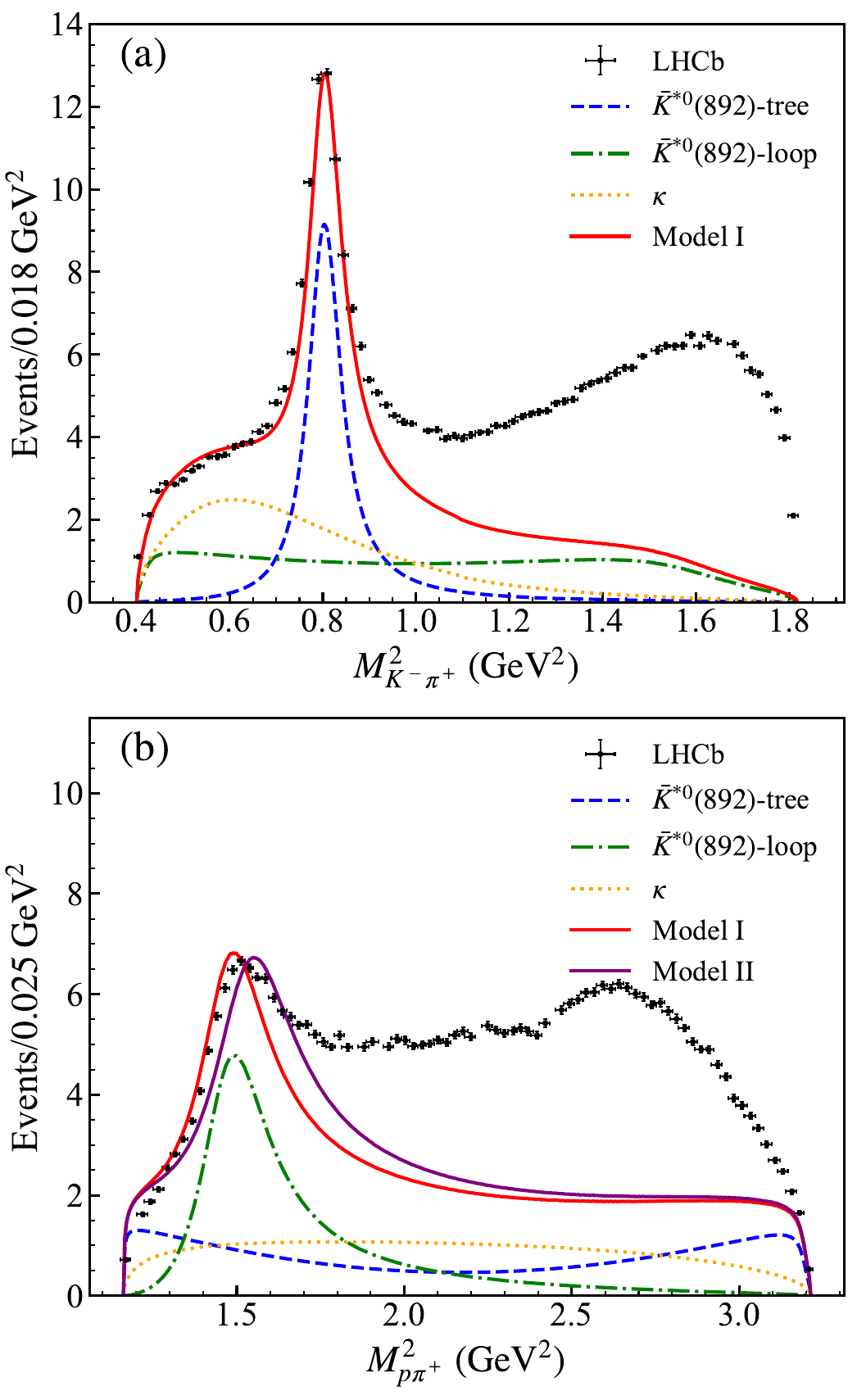}
    \caption{(Color online) The invariant mass distributions for the $\Lambda^+_c \to K^- \pi^+ p$ decay: (a) $M^2_{K^-\pi^+}$ distribution; (b) $M^2_{p\pi^+}$ distribution. Curves labeled as ``$\bar{K}^{*0}(892)$-tree'', ``$\bar{K}^{*0}(892)$-loop'', and ``$\kappa$'' show the results obtained with ${\mathcal{M}}_\mathrm{a}$, ${\mathcal{M}}_\mathrm{b}$, and ${\mathcal{M}}_\kappa$, respectively. Curves labeled as ``Model~I'' and ``Model~II'' represent the total contributions of Eq.~(\ref{eq:amp_total}) obtained with the pole and Breit-Wigner masses and widths of the $\Delta(1232)^{++}$, respectively. The experimental data are taken from Ref.~\cite{LHCb:2022sck} measured by the LHCb Collaboration.}
    \label{fig:imd}
\end{figure}

With the parameters listed in Table~\ref{table:Parameters}, along with the obtained global normalization factor of $C=4\times 10^{13}$ and $V_p = 4\times 10^{-5}$ to match the selected candidates in experiment, we evaluate the $K^- \pi^+$ and $p\pi^+$ invariant mass distributions for the $\Lambda^+_c \to K^- \pi^+ p$ decay in both Model~I and Model~II. In Fig.~\ref{fig:imd}, we present the theoretical results compared with the experimental data~\cite{LHCb:2022sck}. The figure shows the individual contributions from the tree-level $\bar K^{*0}(892)$ diagram, the triangle loop diagram associated with the $\Delta(1232)^{++}$, the dynamically generated $\kappa$ state, and the total contributions for both Model~I and Model~II. In our calculations, the cutoff parameter is determined as $\Lambda=1.3$~GeV for Model~I and $\Lambda=1.5$~GeV for Model~II, respectively.

As illustrated in Fig.~\ref{fig:imd}(a), the prominent $\bar K^{*0}(892)$ peak in the $K^- \pi^+$ invariant mass spectrum is successfully reproduced by the tree-level amplitude. Additionally, the $\bar K^{*0}(892)$ loop amplitude provides an almost constant background across the $K^-\pi^+$ spectrum. However, these contributions combined still fail to account for the event yield in the low-energy region. As indicated by the orange dotted curve in Fig.~\ref{fig:imd} (a), this missing yield is successfully filled by the $\kappa$ resonance, which is dynamically generated from the $S$-wave $K\pi$ final-state interaction.

Turning to the $p \pi^+$ invariant mass distribution in Fig.~\ref{fig:imd} (b), the $\pi^+ p$ final-state rescattering mechanism, driven by the aforementioned tree-level $\bar K^{*0}(892)$ diagram, naturally generates the $\Delta(1232)^{++}$ peak.~\footnote{The theoretical results for the $p \pi^+$ invariant mass distributions are scaled by an extra factor of $25/18$ to match the different bin widths of the experimental data: $0.018~\mathrm{GeV}^2$ for the $K^-\pi^+$ invariant mass distributions and $0.025~\mathrm{GeV}^2$ for the $p \pi^+$ invariant mass distributions.} A detailed comparison reveals a visible difference between the two parameterizations: Model~I, which utilizes the pole mass and width (red solid curve), excellently reproduces the experimental data. In contrast, Model~II, based on the Breit-Wigner parameters (purple solid curve), yields a broader distribution with its peak deviating from the experimental position toward a higher invariant mass. The observation that the pole parameterization provides a better description of the LHCb measurements than the conventional Breit-Wigner form indicates that the $\Delta(1232)$ in the $\Lambda_c^+ \to K^- \pi^+ p$ decay is more likely to dynamically originate from the $\pi^+ p$ rescattering process.

\begin{figure}[htbp]
    \centering
    \includegraphics[scale=0.45]{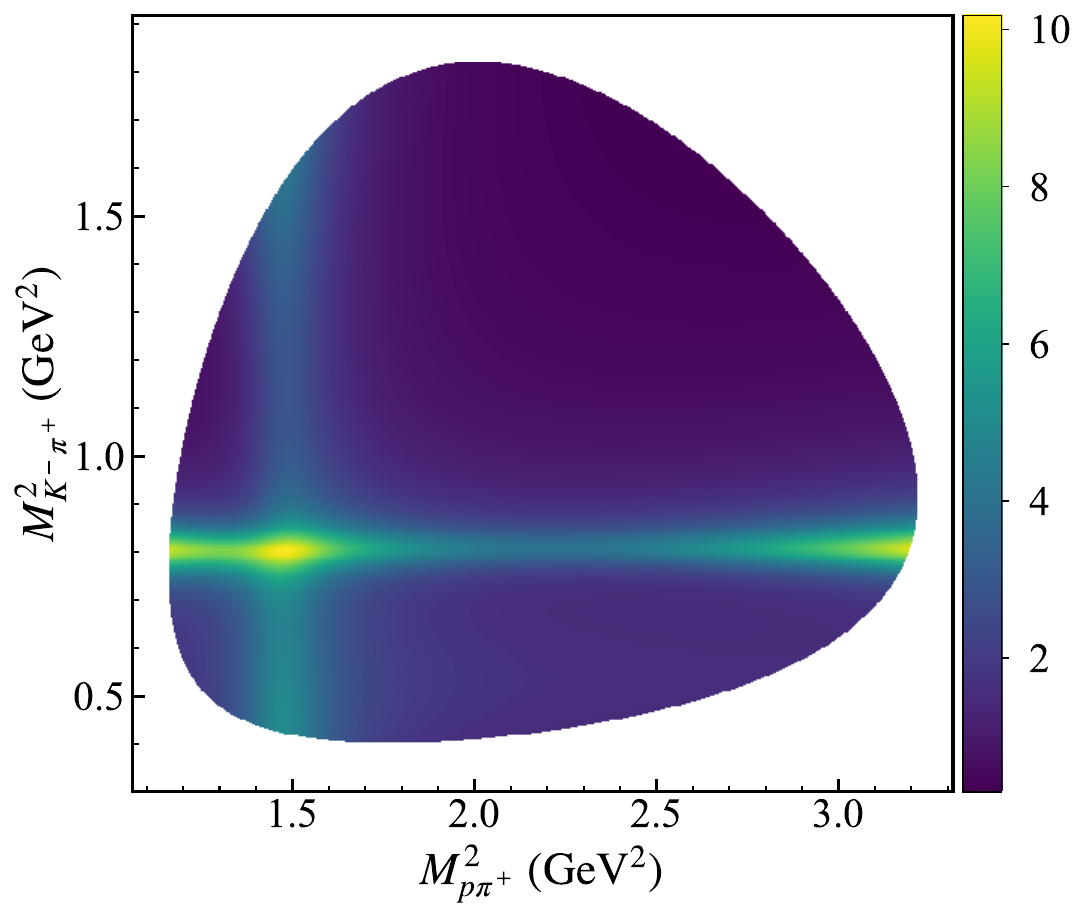}
    \caption{Dalitz plot for the invariant masses of $K^- \pi^+$ and $p \pi^+$ for the $\Lambda^+_c \to K^- \pi^+ p$ decay.}
    \label{fig:Dalitz}
\end{figure}

Furthermore, to visualize the phase space distribution generated by our amplitude model, we present the theoretical Dalitz plot for the $\Lambda^+_c \to K^- \pi^+ p$ decay in Fig.~\ref{fig:Dalitz}. In line with the production mechanisms discussed in this work, the plot displays distinct resonance bands corresponding exclusively to the $\bar K^{*0}(892)$ and $\Delta(1232)^{++}$ states. It can be seen from Fig.~\ref{fig:Dalitz} that the kinematic production region for low-lying excited $\Lambda$ states (for example, $\Lambda(1520)$ and $\Lambda(1670)$) in the $K^- p$ system is well separated from those of $\bar{K}^{*0}$ and $\Delta^{++}$. Their contributions can therefore be safely ignored. Even if they do have some contributions, they will contribute as a background and leave the line shapes of the $K^- \pi^+$ and $p \pi^+$ invariant mass distributions unchanged.
 
Finally, to quantify the yield of the $\Delta(1232)^{++}$ resonance, we evaluate the production ratio of $\Delta^{++}$ resonance in the $\Lambda_c^+ \to K^- \Delta(1232)^{++}$ process with the intermediate triangle loop diagrams as shown in Fig.~\ref{fig:Feynman}(b). The corresponding partial decay width is given by
\begin{align}\label{eq:Del_Gamma}
    \Gamma_{\Lambda_c^+ \to \Delta(1232)^{++} K^-}= \frac{|\vec{p}_4|}{16 \pi M_{\Lambda_c^+}^2} |\mathcal{M}'|^2\,,
\end{align}
where $|\vec{p}_4|$ is the momentum of $\Delta(1232)^{++}$ (or $K^-$) in the $\Lambda_c^+$ rest frame, and the decay amplitude $\mathcal{M}'$ is expressed as
\begin{align}\label{eq:amp_Del}
    \mathcal{M}'&= \frac{g_{\Delta^{++} \pi^+ p} g_{\Lambda_c^+ p \bar K^{*0}} g_{\bar K^{*0} \pi^+ K}}{m_3}\bar{u}^\mu (p_4)\int \frac{\mathrm{d}^4 q_1}{(2\pi)^4} q_{3\mu} \nonumber\\
    &\times G_{1/2}(q_2,m_2,\epsilon) ( \gamma_5+1) \gamma_\alpha G^{\alpha\beta}_1(q_1,m_1, \Gamma_{\bar{K}^{*0}}) \nonumber\\
    &\times  (p_3-q_3)_\beta G_0(q_3,m_3,\epsilon) u(p_0)F(q_3^2,m_3^2) \,.
\end{align}
Using Eqs.~(\ref{eq:g_LamcK892p}) and (\ref{eq:Del_Gamma}), we can deduce the branching fraction ratio of the $\Lambda_c^+\to \Delta(1232)^{++} K^-$ decay to the $\Lambda_c^+\to p\bar{K}^{*0}(892)$ decay. For Model~I, we obtain
\begin{align}\label{eq:R_I}
    R^\mathrm{I} = \frac{\mathcal{B}[\Lambda_c^+\to \Delta(1232)^{++} K^-]}{\mathcal{B}[\Lambda_c^+\to p\bar{K}^{*0}(892)]}=0.52\,,
\end{align}
and similarly, for Model~II,
\begin{align}\label{eq:R_II}
    R^\mathrm{II} = \frac{\mathcal{B}[\Lambda_c^+\to \Delta(1232)^{++} K^-]}{\mathcal{B}[\Lambda_c^+\to p\bar{K}^{*0}(892)]}=0.54\,.
\end{align}
For comparison, the corresponding experimental ratio is determined to be~\cite{ParticleDataGroup:2024cfk}
\begin{align}\label{eq:R_exp}
    R^\text{exp} = \frac{\mathcal{B}[\Lambda_c^+\to \Delta(1232)^{++} K^-]}{\mathcal{B}[\Lambda_c^+\to p\bar{K}^{*0}(892)]}=1.27\pm0.09\,.
\end{align}

As seen from the above results, the theoretical predictions ($R \approx 0.5$) are noticeably smaller than the experimental central value ($R^\text{exp} \approx 1.27$). This discrepancy can be largely attributed to the complex interference patterns inherent in the three-body decay dynamics. In the experimental amplitude analysis, the isolated contribution of the $\Delta(1232)^{++}$ actually exceeds the overall experimental data points in the peak region, indicating significant destructive interference with other amplitudes to reproduce the measured spectrum~\cite{LHCb:2022sck}. In contrast, our dynamical model favors a constructive interference pattern. Consequently, our model requires a relatively smaller isolated $\Delta(1232)^{++}$ contribution to properly describe the same data, naturally resulting in a smaller theoretical ratio. This is consistent with that the predictions that the $\Lambda^+_c \to \Delta^{++} K^-$ decay would proceed through a color-suppressed $W^+$ exchange diagram ($c+d \to s+u$), which generally gives small contributions~\cite{Cheng:2010vk,Cheng:2015iom,Miyahara:2016yyh,Xie:2014tma}.

\section{Summary}\label{sec:summary}

In this work, we investigated the $\Lambda_c^+ \to K^- \pi^+ p$ decay to elucidate the production mechanisms of the involved resonances, with a particular focus on the production of $\Delta(1232)^{++}$ state in the $\pi^+ p$ system. Within our framework, the tree-level $\Lambda_c^+\to p\bar{K}^{*0}(892)\to K^-\pi^+ p$ decay was evaluated using the effective Lagrangian approach, which properly accounted for the dominant $\bar K^{*0}(892)$ peak. Driven by this initial process, the $\Delta(1232)^{++}$ state was dynamically generated via the subsequent rescattering of $\pi^+ p$ through a triangle loop mechanism. Furthermore, to address the low-energy region of the $K^-\pi^+$ spectrum, we employed the chiral unitary approach to incorporate the $\kappa$ contribution, which is dynamically generated by the $S$-wave $K^-\pi^+$ final-state interaction.

Our numerical results were compared with the LHCb measurements for the process $\Lambda_c^+ \to K^- \pi^+ p$. The tree-level process reproduced the dominant $\bar K^{*0}(892)$ peak in the $K^-\pi^+$ invariant mass spectrum, and the $\Delta(1232)^{++}$ peak was also captured by the $\pi^+ p$ rescattering. To further investigate the nature of the $\Delta(1232)^{++}$, we tested two parameterizations: the pole parameters (Model~I) and the conventional Breit-Wigner form (Model~II). The calculation showed that Model~I yielded a better description of the experimental data. This implies that the $\pi^+ p$ final-state rescattering plays a substantial role in the $\Delta(1232)^{++}$ production in this specific decay.

Finally, we calculated the branching fraction ratio $\mathcal{B}[\Lambda_c^+\to \Delta(1232)^{++} K^-]/\mathcal{B}[\Lambda_c^+\to p\bar{K}^{*0}(892)]$. The theoretical prediction ($\sim 0.5$) is smaller than the experimental measurement ($1.27\pm0.09$). A possible explanation for this difference is that the destructive interference in the experimental amplitude analysis leads to a larger isolated $\Delta(1232)^{++}$ fraction, whereas our dynamical model features constructive interference. Future experimental measurements at BESIII, Belle II, and LHCb are expected to clarify these interference effects and test our model calculations.

\begin{acknowledgements}\label{sec:acknowledgements}

This work is partly supported by the National Key R\&D Program of China under Grant Nos. 2023YFA1606703 and 2024YFE0105200; by the National Natural Science Foundation of China under Grant Nos. 12575094, 12575081, 12435007, 12361141819, and 12475086; and by Taishan Scholar Project of Shandong Province.

\end{acknowledgements}

\bibliography{Refs.bib}

\end{document}